\documentclass[twocolumn,3p,times]{elsarticle}
  
\usepackage{mathptmx, courier, pifont}
\usepackage[scaled=0.92]{helvet}
\usepackage[T1]{fontenc}
\usepackage{textcomp}
\usepackage{amsfonts,mathastext}

\usepackage{epsfig}
\usepackage{amssymb}
\usepackage{graphicx}
\usepackage{amssymb}
\usepackage{graphicx}
\usepackage{mathtools, cuted}
\biboptions{numbers,sort&compress}
\usepackage{braket}
\usepackage{physics}

\begin{document}

\begin{frontmatter}

\title{Enhanced symmetry energy bears universality of the r-process}

\author[a,b]{Jos\'e Nicol\'as Orce\corref{mycorrespondingauthor}}
\cortext[mycorrespondingauthor]{Corresponding author}
\ead{jnorce@uwc.ac.za}
\ead[url]{https://nuclear.uwc.ac.za}

\author[c]{Balaram Dey}

\author[a]{Cebo Ngwetsheni}

\author[d]{Srijit Bhattacharya} 

\author[e,f]{Deepak Pandit} 

\author[a]{Brenden  Lesch}

\author[a]{Andile  Zulu}

\address[a]{Department of Physics \& Astronomy, University of the Western Cape, P/B X17, Bellville ZA-7535, South Africa}

\address[b]{National Institute for Theoretical and Computational Sciences (NITheCS), South Africa}

\address[c]{Department of Physics, Bankura University, Bankura-722155, West Bengal, India}

\address[d]{Department of Physics, Barasat Govt. College, Barasat, West Bengal-700124, India}

\address[e]{Variable Energy Cyclotron Centre, 1/AF-Bidhannagar, Kolkata-700064, India}
\address[f]{Homi Bhabha National Institute, Training School Complex, Anushaktinagar, Mumbai 400094, India}

\date{\today}

\begin{abstract}

The abundance of about half of the stable nuclei heavier than iron via the rapid neutron capture process or $r$-process is intimately related to the competition between neutron capture and $\beta$-decay rates, which ultimately depends on the binding energy of neutron-rich nuclei. 
The well-known Bethe-Weizs\"acker semi-empirical mass formula\cite{weiz,bethe}  describes the binding energy of ground states --  i.e. nuclei with temperatures of 
$T\approx0$ MeV --  with the symmetry energy parameter converging between $23-27$ MeV for heavy nuclei. 
Here we find an unexpected enhancement of the symmetry energy at higher temperatures, $T\approx0.7-1.0$ MeV, from the available data of giant dipole resonances built on excited states. Although these are likely the temperatures where seed elements are created -- during the cooling down of the ejecta following neutron-star mergers\cite{mergersnucleo}  or collapsars\cite{collapsar} --  the fact that the symmetry energy remains constant between $T\approx0.7-1.0$ MeV, suggests a similar trend down to $T\approx0.5$ MeV, where neutron-capture may start occurring. Calculations using this relatively larger symmetry energy yield a reduction of the binding energy per nucleon for heavy neutron-rich nuclei and inhibits radiative neutron-capture rates. This results in a substantial close in of the neutron dripline -- where nuclei become unbound -- which elucidates the 
long sought universality of heavy-element abundances through the $r$-process; as inferred from the similar abundances found in extremely metal-poor stars and the Sun.

\end{abstract}


\end{frontmatter}

The binding energy of a nucleus with $Z$ protons and $N$ neutrons can be described by the well-known Bethe-Weizs\"acker semi-empirical  
mass formula (SEMF) \cite{weiz,bethe}, 
\begin{eqnarray}
 B(Z,A)&=&a_vA-a_sA^{2/3}-a_cZ(Z-1)A^{-1/3} \nonumber \\
 &-&a_{sym}\frac{(A-2Z)^2}{A} \pm a_pA^{-3/4}, 
 \label{eq:bw}
\end{eqnarray}
where  $A=Z+N$ is the mass number and $a_v, a_s, a_c, a_{sym}$ and $a_p$ are the volume, surface, Coulomb, symmetry energy and pairing coefficients, respectively. 
The symmetry energy, $a_{sym}(A)(N-Z)^2/A$, reduces the  
total binding energy $B(Z,A)$ of a nucleus as the neutron-proton asymmetry becomes larger, i.e. for $N\gg Z$, 
and yields the typical negative slope of the binding energy curve\cite{Krane} 
for $A>62$. 
It is divided by A to reduce its importance for heavy nuclei, and it 
depends on the mass dependency of  $a_{sym}(A)$. Its convergence 
for  heavy nuclei  establishes the frontiers of the neutron dripline for particle-unbound nuclei 
and eventually leads to the disappearance of protons at extreme nuclear densities\cite{marek1}. 

Furthermore, $a_{sym}(A)$ is relevant 
to understanding  neutron skins\cite{nskin2}, the effect of three-nucleon forces\cite{hebeler} and -- 
through the equation of state (EoS) --  supernovae cores, neutron  stars and binary mergers\cite{neutronstars,latimer,pearson}. 
The latter are the first known astrophysical site where 
heavy elements are created through the rapid neutron-capture or $r$-process\cite{rprocess1,rprocess}. 
The identification of heavy elements in neutron star mergers is supported by the short
duration gamma-ray bursts via their infrared afterglow\cite{kilonova} -- only understood
by the opacities of heavy  nuclei -- as well as blueshifted Sr II absorption lines\cite{sr}, following the   
expansion speed of the ejecta gas at $v=0.1-0.3~c$.
Mergers are 
expected to be the only source for the creation of elements above lead and bismuth, as inferred from the very scarce abundance of actinides 
in the solar system\cite{actinides}. 

The universality of the $r$-process for the  heaviest elements with $56 < Z < 90$ is further inferred from the similar abundance patterns  
observed in both extremely metal-poor stars and the Sun\cite{christlieb,sneden}. 
Other potential sources of heavy elements involve different types of supernova (e.g. collapsars~\cite{collapsar} -- the supernova-triggering 
collapse of rapidly  rotating massive stars -- and type-II supernova\cite{rprocess1}), which need to be considered in order 
to explain all neutron-capture abundances\cite{qian,aoki2}.

It is the motivation of this work to understanding the limits of the neutron dripline and heavy-element production by investigating $a_{sym}(A)$ at different temperatures $T$ using available data of potential interest to the $r$-process; namely, data from photoabsorption cross sections, binding energies and giant dipole resonances. 

Generally, $a_{sym}(A)$ is  parametrized using the
leptodermous approximation  of Myers and Swiatecki, where $A^{-1/3} \ll
1$\cite{myers}, 
\begin{equation}
	a_{sym}(A)=S_v\left(1-\frac{S_s}{S_v}A^{-1/3}\right),
	\label{eq:asym}
\end{equation}
which considers the  modification of the volume
symmetry energy, $S_v$, by the surface symmetry energy $S_s$. This particular leptodermous parametrization was chosen on the account of 
its better fit to the masses of isobaric nuclei~\cite{tian}. Constraints on these parameters have been investigated using experimental and theoretical information concerning properties of ground states, i.e. at $T=0$ MeV\cite{lattimer,trippa}.

The giant dipole resonance ({\small GDR}) represents the  main contribution to the absorption and emission of 
electromagnetic radiation (photons) in nuclei\cite{berman1975}. The dynamics of this quantum collective excitation 
is characterized by the inter-penetrating motion of proton and neutron fluids out of phase\cite{migdal}, 
which results from the density-dependent symmetry energy, $a_{sym}(A)(\rho_{_N}-\rho_{_Z})^2/\rho$, 
acting as a restoring force\cite{berman1975}; where $\rho_{_N}$, $\rho_{_Z}$ and $\rho=\rho_{_N}+\rho_{_Z}$ are the neutron, 
proton and total density, respectively, which spread  uniformly throughout the nucleus.

The ratio of the induced dipole moment to an applied constant electric field 
yields the static nuclear polarizability, $\alpha$. 
Using the hydrodynamic model and assuming  inter-penetrating 
proton and neutron fluids with a well-defined nuclear surface of radius $R=r_{_0}A^{1/3}$ fm 
and  $\rho_{_Z}$ as the potential energy of the liquid drop, 
Migdal\cite{migdal} obtains the following relation between the static nuclear polarizability, $\alpha$, 
and $a_{sym}$, 
\begin{equation}
\alpha=\frac{e^2R^2A}{40 a_{sym}}=2.25\times 10^{-3} A^{5/3} \mbox{fm}^3, 
\label{eq:sigma-2}
\end{equation}
where $r_{_0}=1.2$ fm, $e^2=1.44$ MeV fm in the c.g.s. system, and a constant value of 
$a_{sym}=23$ MeV was utilized.

Alternatively,  $\alpha$  can be calculated for the ground states of nuclei 
using second-order perturbation theory\cite{levinger2} following the sum rule,
\begin{eqnarray}
	\alpha&=&2e^2\sum_n \frac{\langle i\parallel\hat{E1}\parallel n\rangle \langle n\parallel\hat{E1}\parallel i\rangle}{E_{_{\gamma}}} \label{eq:polar} 
	\\ 
	&=&\frac{e^2\hbar^2}{M}\sum_n \frac{f_{in}}{E_{_{\gamma}}^2} = \frac{\hbar c}{2\pi^2}\int_0^\infty 
	\frac{\sigma_{_{total}}(E_{_{\gamma}})}{E_{_{\gamma}}^2} ~dE_{_{\gamma}} \label{eq:fsigma} \\ 
	&=& \frac{\hbar c}{2\pi^2}\sigma_{_{-2}},
	\label{eq:sigma-22}
\end{eqnarray}
where $E_{_{\gamma}}$ is the $\gamma$-ray energy corresponding to a transition connecting the ground state $|i\rangle$ 
and an excited state $|n\rangle$, $M$  the nucleon mass, $f_{in}$ the dimensionless 
oscillator strength for $E1$ transitions~\cite{levinger2}
and $\sigma_{_{-2}}$ the second moment of the total electric-dipole 
photo-absorption cross section, 
\begin{eqnarray}
	\sigma_{_{-2}}&=&\int_0^\infty\frac{\sigma_{_{total}}(E_{_{\gamma}})}{E_{_{\gamma}}^2} ~dE_{_{\gamma}},
	\label{eq:sigma-222}
\end{eqnarray}
where $\sigma_{_{total}}(E_{_\gamma})$ is the total photo-absorption cross section, which generally includes 
$(\gamma,n) + (\gamma,pn) + (\gamma,2n) + (\gamma,3n)$ photoneutron and  available photoproton and photofission cross sections~\cite{kawano},   
in competition in the  {\small GDR} region \cite{GDRreview,lectures}. 
By comparing Eqs. \ref{eq:sigma-2} and \ref{eq:sigma-22}, a mass-dependent symmetry energy, $a_{sym}$(A), is extracted
in units of MeV,
\begin{eqnarray}
	a_{sym}(A)&=&\frac{e^2R^2\pi^2A}{20~\hbar c~\sigma_{_{-2}}}\label{eq:asym1}
	\approx 5.2\times10^{-3}~\frac{A^{5/3}}{\sigma_{_{-2}}}.
\end{eqnarray}
Empirical evaluations reveal that $\sigma_{_{-2}}$ can also be approximated by $\sigma_{_{-2}} = 2.4\kappa A^{5/3}$, 
where the dipole polarizability parameter $\kappa$ measures {\small GDR} deviations between experimental and hydrodynamic model predictions\cite{orce_review}.  

\begin{figure}[!h]
	\begin{center}
		\includegraphics[width=7.5cm,height=5.5cm,angle=-0]{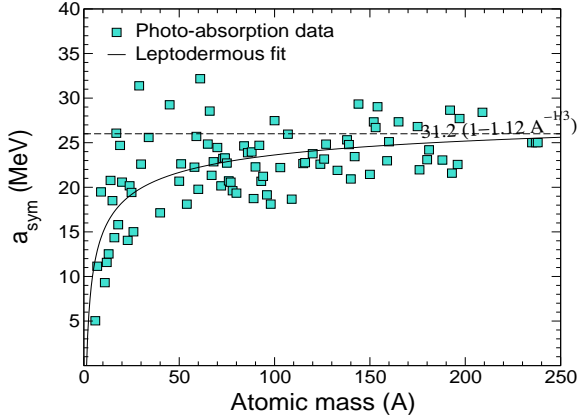} 
		\caption{Symmetry energy coefficient, $a_{sym}(A)$, of finite nuclei  as a function of mass number $A$ extracted from the experimental 
		$\sigma_{_{-2}}$ values extracted from available photoabsorption cross-sections\cite{exfor,ENDF}, as given in Eq. \ref{eq:asym1} and fitted (solid line) by Eq.~\ref{eq:asym}.}
		\label{fig:sigma_-2}
	\end{center}
\end{figure}

Figure~\ref{fig:sigma_-2} shows the distribution of $a_{sym}(A)$ for the ground state of stable isotopes, along the nuclear landscape,  
determined from empirical $\sigma_{_{-2}}$ values. Data include all available emission channels. 
The contribution of $(\gamma,p)$ cross sections are evident in light nuclei, 
which significantly reduces the symmetry energy. For heavy nuclei, $(\gamma,n)$ cross sections are dominant because 
of the higher Coulomb barrier.  A fit to the data using Eq.~\ref{eq:asym} (solid line) yields $a_{sym}(A)=31.2(12)\left(1-1.12(10)A^{-1/3}\right)$ MeV, with an {\small RMS} deviation of 22\%\cite{orce2}. Unfortunately, $(\gamma,p)$ cross-section data are very scarce, which 
directly affects the $a_{sym}(A)$ trend for $A\lessapprox70$ in Fig.~\ref{fig:sigma_-2}.

In addition, Tian and co-workers determined $a_{sym}(A)=28.32\left(1-1.27A^{-1/3}\right)$ MeV from a global fit to the binding energies of isobaric nuclei with mass number $A\geq10$\cite{tian} -- extracted from the 2012 atomic mass evaluation\cite{audi} --  
with $S_v\approx 28.32$ MeV being the bulk symmetry energy coefficient and $\frac{S_s}{S_v}\approx 1.27$ the surface-to-volume ratio. 
Similar coefficients are calculated in Refs.\cite{neutronstars,Danielewicz}.
Within this approach,  the extraction of $a_{sym}(A)$ only depends on the Coulomb energy term in the SEMF  
and shell effects\cite{massmodel} --  which are both included~\cite{tian} -- and $a_{sym}(A)$ presents a maximum energy around 23 MeV.
This description of $a_{sym}(A)$ has been used to explain the enhanced $\sigma_{_{-2}}$ values observed for low-mass nuclei\cite{orce2}.

The symmetry energy $a_{sym}(A)$ is the fundamental parameter that characterizes the energy of the {\small GDR}, $E_{_{_{GDR}}}$, 
within the Steinwedel-Jensen ({\small SJ}) model of proton and neutron compressible fluids moving within the rigid surface of the nucleus\cite{steinwedel}.  Danos improved the {\small SJ} model by including  the {\small GDR} width, $\Gamma_{_{GDR}}$\cite{danos2,berman1975} 
in the second-sound hydrodynamic model\cite{danos2,berman1975}, where  $E_{_{_{GDR}}}$ and $\Gamma_{_{GDR}}$ are related to $a_{sym}(A)$ as\cite{lectures}, 
\begin{eqnarray}
 a_{sym}(A) &=& {\frac{M A^2}{8\hbar^2K^2 NZ}   \frac{ E_{_{_{GDR}}}^2}{1-\bigg(\frac{\Gamma_{GDR}}{2 E_{_{_{GDR}}}}\bigg)^2}}  \nonumber \\
 &\approx& {1 \times 10^{-3} \bigg(\frac{A^{8/3}}{NZ}\bigg) \frac{ E_{_{_{GDR}}}^2}{1-\bigg(\frac{\Gamma_{GDR}}{2 E_{_{_{GDR}}}}\bigg)^2}},
\label{GDRsymm}
\end{eqnarray}
where  $K$ is the real eigenvalue of $\bigtriangledown^2\rho_{_Z}+K^2\rho_{_Z}=0$, with the 
boundary condition $(\mathbf{\hat{n}} \bigtriangledown\rho_{_Z})_{_{surface}}=0$, and has a value of 
$KR=2.082$ for a spherical nucleus\cite{rayleigh}.
For quadrupole deformed nuclei with an eccentricity of $a^2-b^2=\epsilon R^2$, where 
$a$ and $b$ are the half axes and $\epsilon$ the deformation parameter, the {\small GDR} lineshape splits into 
two peaks with similar values of $Ka$ and $Kb\approx2.08$~\cite{danos2}. 
For deformed nuclei, we estimate a similar equation to Eq. \ref{GDRsymm}, but using the average centroid energy and the 
{\small FWHM} of the total Lorentzian (see e.g.\cite{158Er}). Uncertainties in the quoted values arise from the error propagation of Eq.~\ref{GDRsymm}.

\begin{figure}[!ht]
	\begin{center}
		\includegraphics[width=6.cm,height=4.5cm,angle=-0]{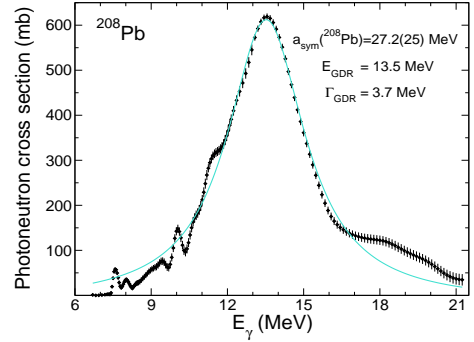} 
		\caption{A Lorentzian fit to {\small GDR} data extracted for $^{208}$Pb~\cite{exfor,ENDF}.}
		\label{fig:lfit_208pb}
	\end{center}
\end{figure}

\begin{figure*}[!ht]
	\begin{center}
				\includegraphics[width=13.5cm,height=5.5cm,angle=-0]{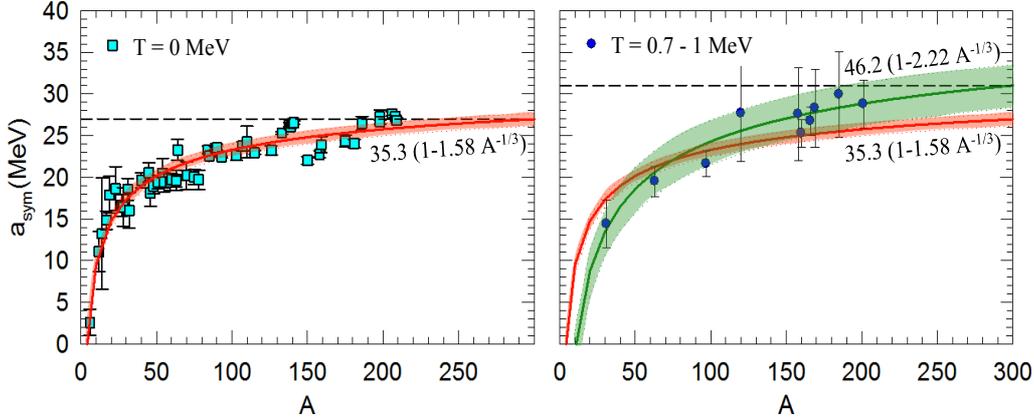}
		\caption{Symmetry energy coefficient, $a_{sym}(A)$, of finite nuclei  as a function of mass number 
		$A$ extracted  from {\small GDRs} built on ground states ($T=0$) (left panel) 
		and  excited states ($T=0.7-1$ MeV) (right panel) using Eq.~\ref{GDRsymm}.}
		\label{fig:gdr_symeng}
	\end{center}
\end{figure*}

The {\small GDR} cross-section data for each nucleus were obtained from the {\small EXFOR} and {\small ENDF} databases~\cite{exfor,ENDF} and fitted with one or two 
Lorentzian curves to extract $E_{_{_{GDR}}}$ and $\Gamma_{_{GDR}}$, as shown e.g. in Fig.~\ref{fig:lfit_208pb} for $^{208}$Pb. 
The data set for each nucleus was selected based on the number of data points, experimental method and energy range. 
In this work, the maximum integrated $\gamma$-ray energy, E$_\gamma ^{max}$, was in the range 20--50 MeV, 
therefore excluding contributions resulting from high energy effects such as pion exchange and other meson resonances. 
The resulting distribution of $a_{sym}(A)$ is shown in the left panel of Fig.~\ref{fig:gdr_symeng}, which converges 
at approximately 27 MeV for heavy nuclei. 
It is reassuring that the two methods based on photoabsorption cross-section data --- namely $a_{sym}(A)$ 
extracted from $\sigma_{_{-2}}$ values and parameters of {\small GDR}s built on ground states -- present similar trends.

Data obtained from {\small GDR} parameters at $T=0$ can also be fitted to Eq.~\ref{eq:asym}, which yields 
$a_{sym}(A)=35.3(7)\left(1-1.58(5)A^{-1/3}\right)$ MeV (red solid band in Fig. 3), with an {\small RMS} deviation of 15\%.
Larger values of $S_v=42.8$ and $S_s=89.9$ were determined by Berman using Eq.~\ref{GDRsymm} for 29 nuclei ranging from 
$A=75$ to 209~\cite{berman2}. 
Furthermore, Berman argued that assuming a surface binding energy coefficient of $a_{_S}=20$ MeV in the SEMF, 
the large symmetry to surface energy ratio, $S_s/a_{_S}=4.5$, favors -- as a result of a steeper slope of the binding energy curve for heavy nuclei -- a close-in neutron dripline 
for heavy elements; hence, constraining the reaction network that produces heavy elements by the $r$-process 
in neutron mergers and supernovae. 
Using our value of $S_s=46$ and $a_{_S}=20$ MeV, a more reasonable ratio of $S_s/a_{_S}=2.3$ is determined. 
Slightly smaller values of $a_{_S}\approx17$ MeV are also found in the literature\cite{Krane,Danielewicz}, yielding  
$S_s/a_{_S}=2.7$.

Furthermore, it is interesting to investigate the  behavior of $a_{sym}(A)$ using the available information on 
{\small GDRs} built on excited states, below the critical temperatures and spins where the {\small GDR} width starts broadening; i.e. for moderate average temperatures of $T\lessapprox T_c = 0.7 +37.5/A$ MeV and spins $J$ below the critical angular momentum 
$J \lessapprox J_c=0.6 A^{5/6}$.    
In fact, similar centroid energies,
$E^{exc}_{_{GDR}}$, and resonance  strengths, $S^{exc}_{_{GDR}}$ -- relative to the Thomas--Reiche--Kuhn $E1$ sum rule~\cite{levinger2} -- to those found for the ground-state counterparts~\cite{GDRreview,GDRenergy} 
indicate a common physical origin for all {GDR}s, in concordance with the Brink--Axel hypothesis that  assumes  that a {\small GDR} can be built on
every state in a nucleus~\cite{brink,axel}.

Applying again Eq.~\ref{GDRsymm}, the right panel of Fig.~\ref{fig:gdr_symeng} shows $a_{sym}(A)$ values for  {\small GDRs} built on 
excited states  in slightly-deformed nuclei $^{31}$P~\cite{31P}, $^{63}$Cu~\cite{63Cu}, $^{97}$Tc~\cite{97Tc}, $^{120}$Sn~\cite{120Sn} and $^{201}$Tl~\cite{201Tl}, as well as for well-deformed nuclei in the $A\approx160-180$ mass region ($^{158,160,166}$Er, $^{169}$Tm and $^{185}$Re~\cite{158Er,160166Er,169Tm185Re}). 
With an average temperature between $T\approx0.7$ and 1.0 MeV and below $J_c$, 
these nuclei were selected to investigate the symmetry energy at temperatures relevant to the r-process nucleosynthesis. 
Surprisingly, $a_{sym}(A)$ values for heavy nuclei are relatively larger than previously observed at $T=0$ MeV. 
A fit to the data using Eq.~\ref{eq:asym} (green solid line in Fig.~\ref{fig:gdr_symeng}) yields $a_{sym}(A)=46.2(2.4)\left(1-2.22(14)A^{-1/3}\right)$ MeV, with an {\small RMS} deviation of 6\%. A value of $a_{sym}(A)=30$ MeV for heavy nuclei yields larger values 
of $S_v=44.44$, $S_s=97.32$ and $S_s/a_{_S}=4.87$ (again for $a_{_S}=20$ MeV). 
Two bands showing the loci limits of the two fitting curves at $T=0$ and $T\approx0.7-1$ MeV are shown for comparison. 
Such a distinct behaviour could clearly affect nucleosynthesis 
of heavy elements via the $r$-process during the cooling down of the ejecta.

\begin{figure}[!ht]
	\begin{center}
		\includegraphics[width=6.5cm,height=5cm,angle=-0]{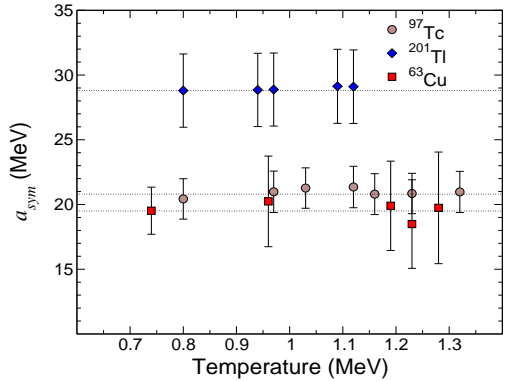} 
		\caption{Symmetry energy coefficient, $a_{sym}(A)$ extracted for $^{63}$Cu, $^{97}$Tc and $^{201}$Tl as a function of temperature, T. 
		A similar constant behaviour is observed for other nuclei. The horizontal dotted lines are shown as a reference.}
		\label{fig:tchange}
	\end{center}
\end{figure}

Lighter or heavier seed nuclei are generally produced depending on the density and temperature of the ejecta gas. 
Assuming nuclear-statistical equilibrium -- when forward and reverse reactions are balanced -- abundances follow a 
Maxwell-Boltzmann distribution where lighter seed nuclei are favoured at very high temperatures ($\propto kT^{-3/2(A-1)}$) and 
heavier nuclei are favoured at very high densities ($\propto \rho^{A-1}$), as those found in the ejecta of neutron-star mergers~\cite{mergersnucleo}.
At temperatures below $T=1$ MeV (or $1.2\times 10^{10}$ K),  seed nuclei are produced before charge reactions freeze out  
-- impeded by the  Coulomb barrier -- at about $T\approx0.5$ MeV  (or $5 \times 10^9$ K). 
Thereafter, heavy nuclei are produced through subsequent neutron capture until 
neutron reactions freeze out -- as neutrons are finally consumed -- at a few 10$^{8}$ K.

Our work may not be sensitive to the lower temperatures occurring 
during neutron capture in neutron-star mergers, which likely range from $T\approx~0.5 \times 10^8$ K~\cite{goriely2011} to 
$T\approx 5 \times 10^9$ K~\cite{wu} (i.e. in the range from $T\approx0.04$ to 0.43 MeV, respectively). 
Nevertheless, Fig.~\ref{fig:tchange} shows that 
the symmetry energy does not change with temperature in the [0.74,1.3] MeV range, which suggests that this 
relation may still hold at lower temperatures.

Such an increase in the symmetry energy results from the change in the effective mass of the nucleon, which decreases as $T$ increases in the temperature interval $0<T<1$ MeV~\cite{donati}.  The temperature dependence of the symmetry energy has been studied within the liquid-drop and Fermi gas models~\cite{bortignon}, where an effective nucleon mass -- the so-called ‘$w$’ mass -- is introduced to account for the non-locality of the Hartree-Fock potential. 
This leads to an increase in the centroid energy of the {\small GDR} and, hence, the symmetry energy of medium and heavy mass nuclei also 
increases by approximately 8\% at $T\approx 1$ MeV~\cite{donati}. 
In the current work, we notice a slight increase of ~ 3-5\% in the centroid energy at $T\approx0.7-1$ MeV as compared with the ground-state values for nearly-spherical $^{120}$Sn~\cite{heck}, $^{208}$Pb~\cite{bau} and $^{201}$Tl~\cite{201Tl} nuclei as well as for the deformed nuclei in the 
$A=160-180$ mass region~\cite{158Er,160166Er,169Tm185Re}.  Although such an increase is within the experimental errors, it leads to a distinct 
systematic behaviour, as shown in the right panel of Fig.~\ref{fig:gdr_symeng}.

\begin{figure*}[!ht]
	\begin{center}
		\includegraphics[width=6.5cm,height=4.5cm,angle=-0]{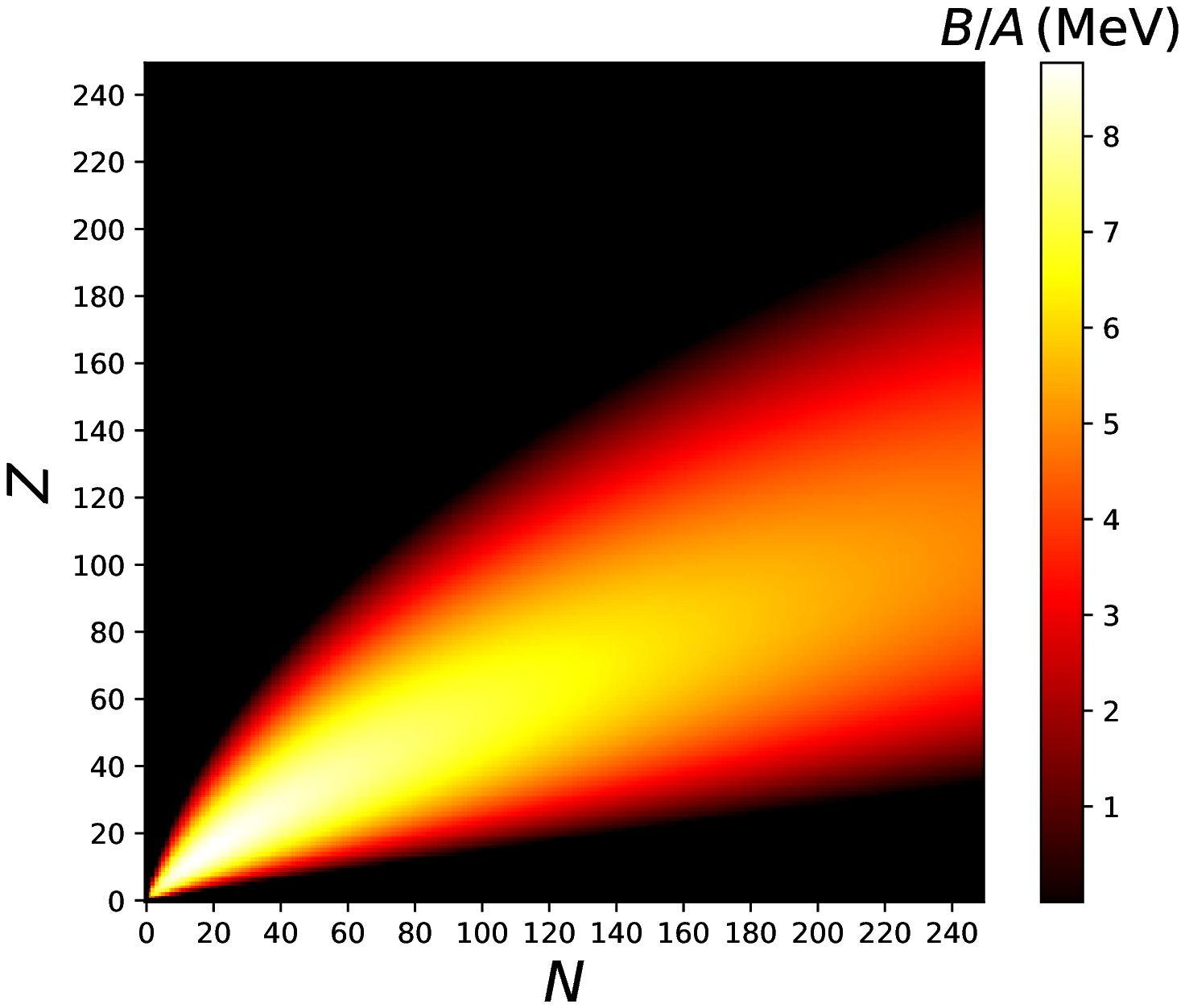}
		\hspace{0.8cm}
				\includegraphics[width=6.5cm,height=4.5cm,angle=-0]{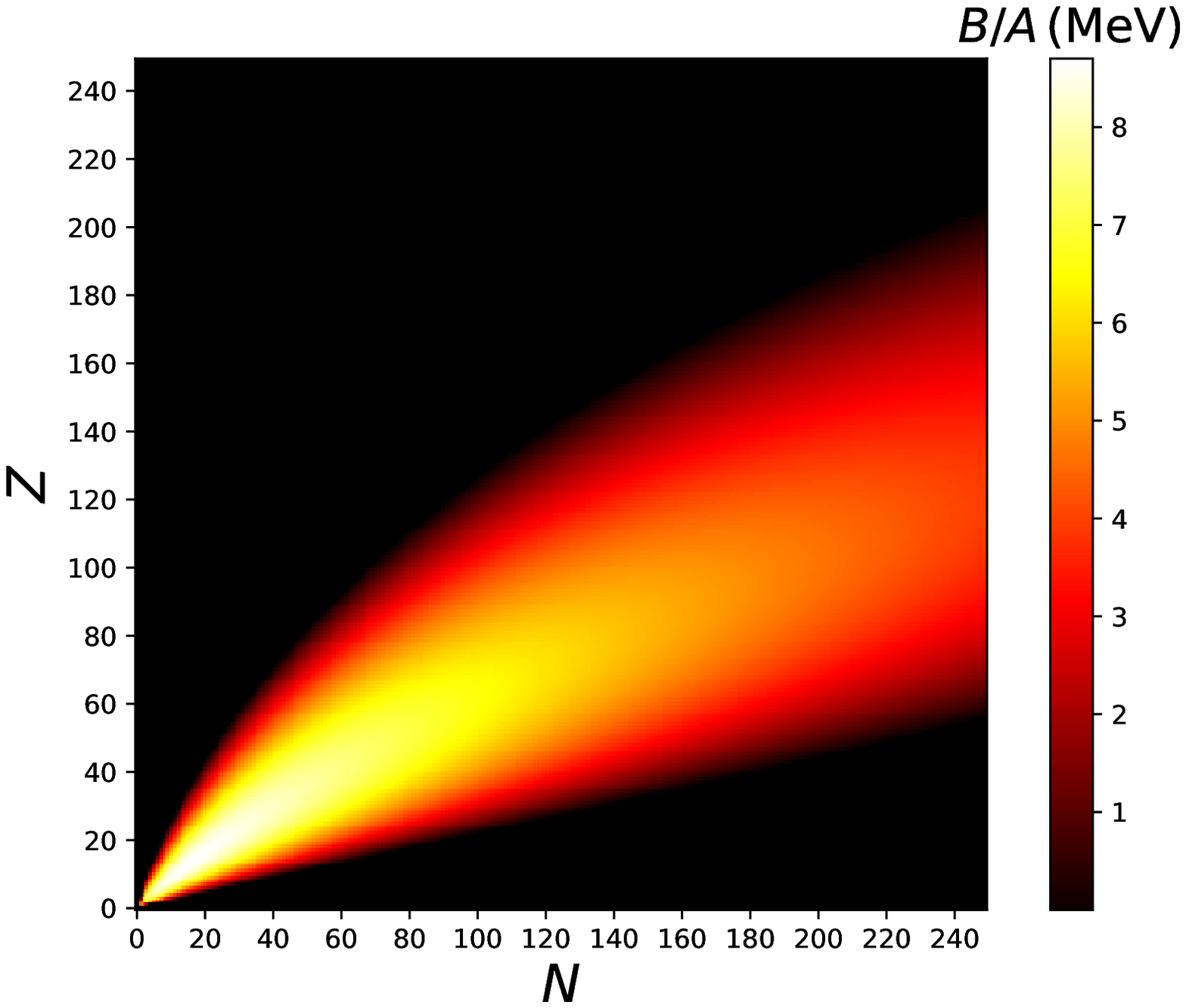}
				\includegraphics[width=6.7cm,height=4.5cm,angle=-0]{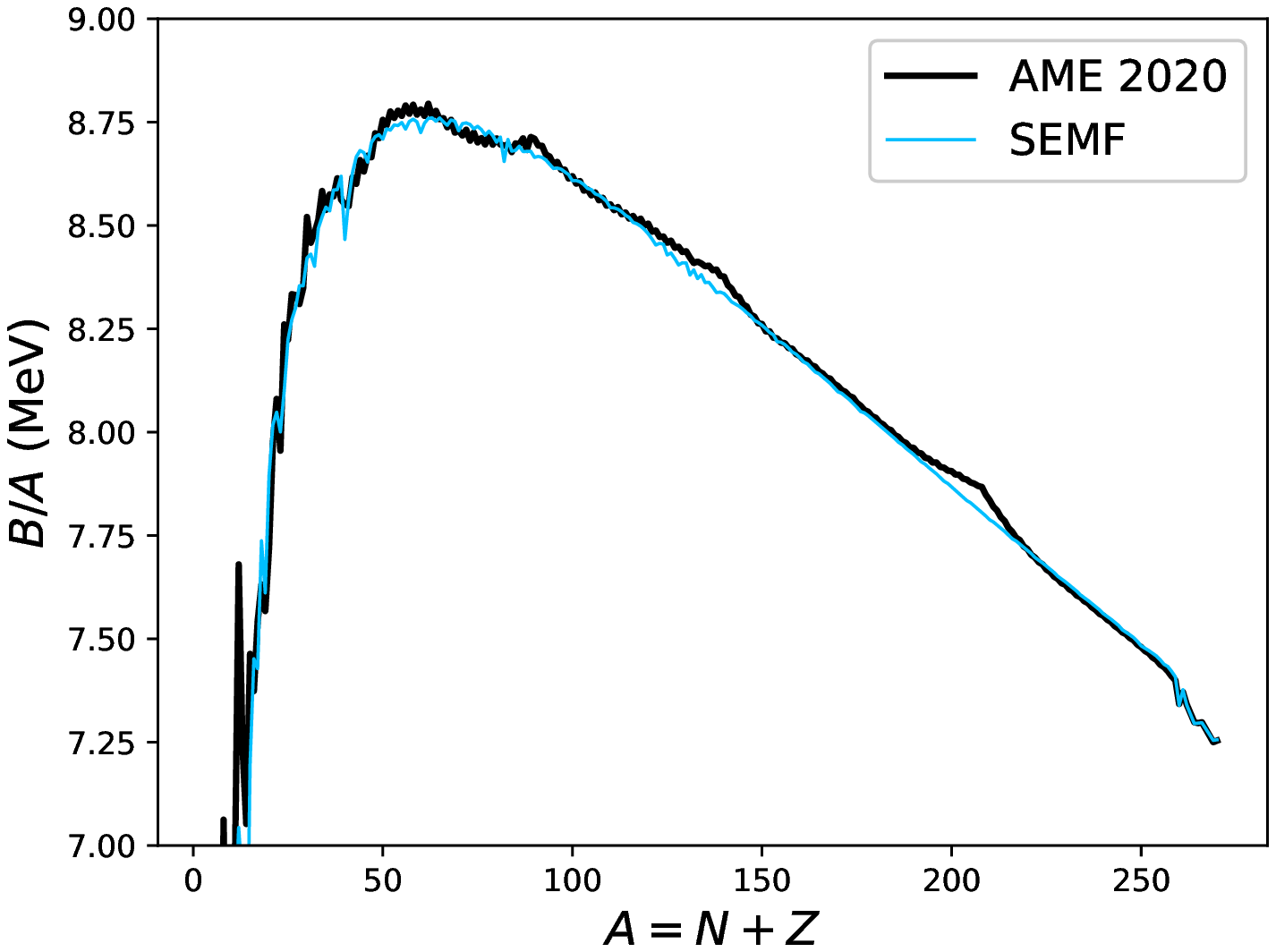}
		\hspace{0.8cm}
				\includegraphics[width=6.7cm,height=4.5cm,angle=-0]{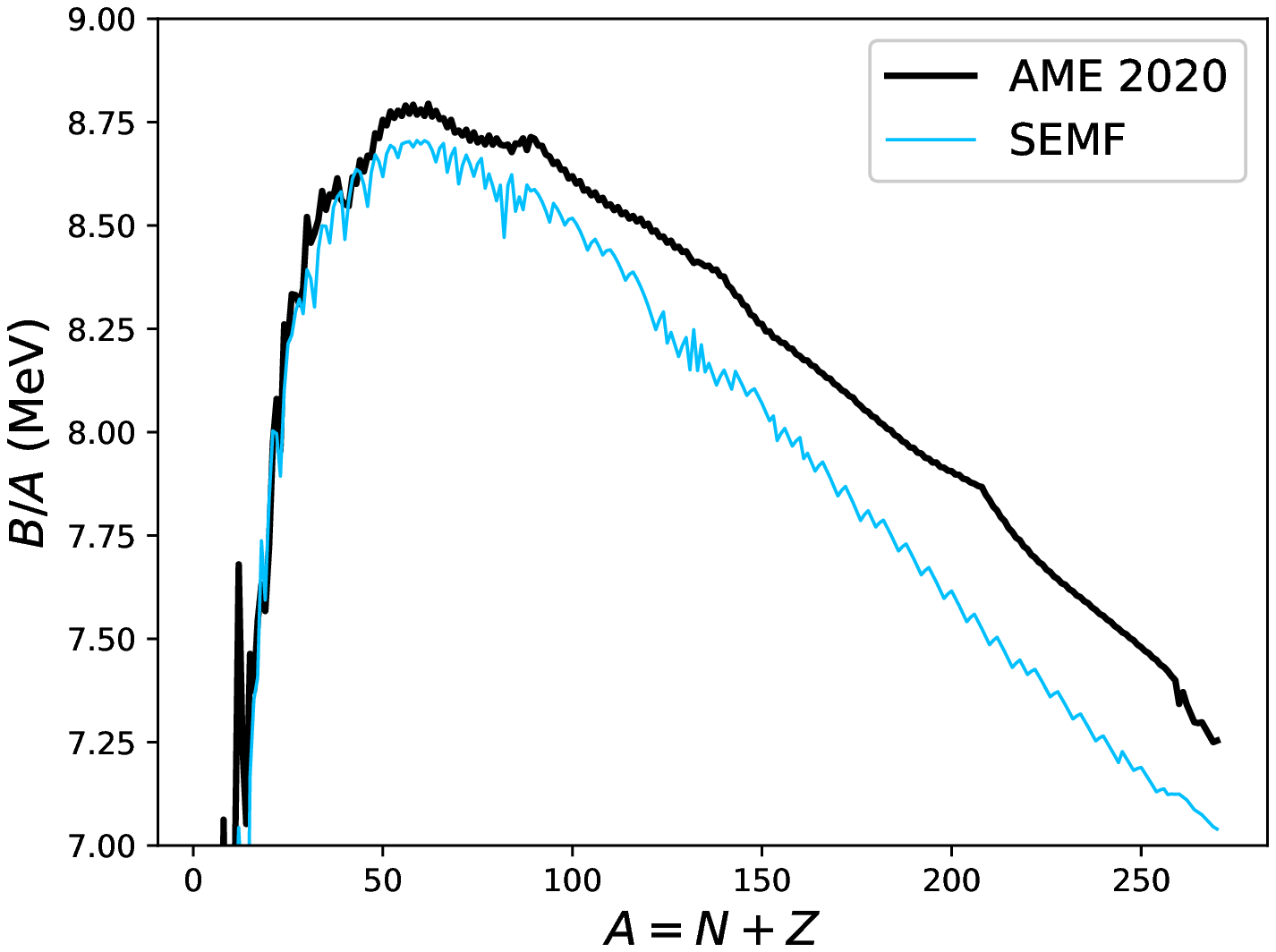}
		\caption{Nuclear charts (top panels) and binding energy curves (bottom panels) showing average 
		binding energies per nucleon using the Bethe-Weizs\"acker SEMF for $a_{sym}=23.7$ \cite{rohlf} MeV (left) and $a_{sym}=30$ MeV (right). 
		Atomic masses in the bottom panels are extracted from the 2020 atomic mass evaluation (AME 2020)~\cite{mass2020}.}
		\label{fig:semf}
	\end{center}
\end{figure*}

\begin{figure}[!ht]
	\begin{center}
		\includegraphics[width=6.5cm,height=5cm,angle=-0]{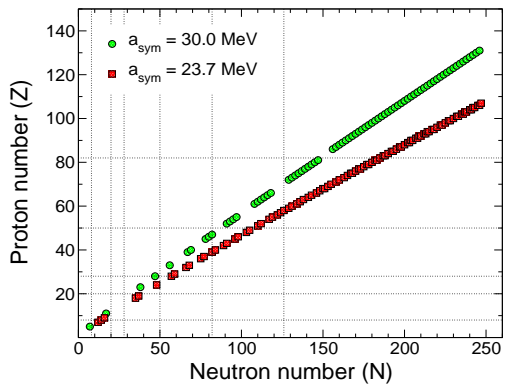} 
		\caption{Neutron driplines predicted at symmetry energy coefficients of $a_{sym}=23.7$~\cite{rohlf} (squares) and $30$ MeV (circles). 
		Dotted lines indicate the proton and neutron magic numbers.}
		\label{fig:dripline}
	\end{center}
\end{figure}

Finally, the effects from the larger symmetry energy at $T\approx0.7-1$ MeV are illustrated in Fig.~\ref{fig:semf}, 
which shows the corresponding nuclear charts (top)  and binding energy curves (bottom) using $a_{sym}=23.7$ MeV\cite{rohlf} (left) and $30$ MeV (right), respectively. 
The nuclear chart determined using $a_{sym}=30$ MeV illustrates a substantial close-in of the neutron dripline, 
as a result of the decreasing binding energy per nucleon in neutron-rich nuclei.   
For instance, the dripline closes in from $^{254}$Pt to $^{220}$Pt for $a_{sym}=23.7$ and $30$ MeV, respectively. 
Figure \ref{fig:dripline} shows the respective neutron driplines and clearly illustrates  the dramatic effect of an 
enhanced symmetry energy in the production of heavy elements, which constrains exotic $r$-process paths and 
plausibly explains the universality of $r$-process abundances inferred from the observation of extremely metal-poor stars and our Sun.

Consequently, such an increase in the symmetry energy leads to the reduction of radiative neutron capture rates as neutron-rich nuclei become less bound. The corresponding change in the capture  cross section has been calculated using {\small TALYS}~\cite{talys} and {\small EMPIRE}~\cite{empire} codes by changing only the mass excess with standard input parameters. Both codes yield similar results with a reduction of the  
neutron-capture cross section by a factor of the order of 10$^2$ in the $A\approx200$ mass region relevant to the $r$-process. 
More detailed calculations will be presented in a separate manuscript.   
These findings support the rapid drop of the neutron capture rates at increasing
neutron excesses inferred from Goriely's microscopic calculations at $T=1.5 \times 10^9$ K~\cite{rprocess1}.

More experimental data regarding {\small GDRs} built on excited states below $T\approx0.7$ MeV are crucially needed in order to elucidate 
the nature of the symmetry energy as a function of temperature. Modern high-efficient spectrometers such as {\small GAMKA} in South Africa~\cite{GAMKA} -- with up to 30 HPGe clover and LaBr$_3$ detectors -- may provide such data.

\end{document}